# Planar topological Hall effect in a uniaxial van der Waals ferromagnet $Fe_3GeTe_2$

Yurong You[1], Yuanyuan Gong[1], Hang Li[2], Zefang Li[2], Mengmei Zhu[1], Jiaxuan Tang[1], Enke Liu[2], Yuan Yao[2], Guizhou Xu[1,a)], Feng Xu[1,a)], Wenhong Wang[2]

[1] School of Materials Science and Engineering, Nanjing University of Science and Technology, Nanjing 210094, China

[2] State Key Laboratory for Magnetism, Beijing National Laboratory for Condensed Matter Physics, Institute of Physics, Chinese Academy of Sciences, Beijing 100190, China

## Abstract

In this work, we reported the observation of a novel planar topological Hall effect (PTHE) in single crystal of $Fe_3GeTe_2$, a paradigmatic two-dimensional ferromagnet with strong uniaxial anisotropy. The Hall effect and magnetoresistance varied periodically when the external magnetic field rotated in the *ac* (or *bc*) plane, while the PTHE emerged and maintained robust with field swept across the hard-magnetized *ab* plane. The PTHE covers the whole temperature region below Tc (~150 K) and a comparatively large value of 2.04 $\mu\Omega$ cm is observed at 100 K. Emergence of an internal gauge field was proposed to explain the origin of this large PTHE, which is either generated by the possible topological domain structure of uniaxial $Fe_3GeTe_2$ or the non-coplanar spin structure formed during the in-plane magnetization. Our results promisingly provide an alternative detection method to the in-plane skyrmion formation and may bring brand-new prospective to magneto-transport studies in condensed matter physics.

---------------------------------------------------------------

a) Authors to whom correspondence should be addressed. E-mail: gzxu@njust.edu.cn, xufeng@njust.edu.cn

Emergent electromagnetism[1] induced by the internal gauge field has triggered many novel and interesting transport phenomena, for instance, the unusual large anomalous Hall effect (AHE) in non-collinear antiferromagnets[2-4] or contributed by topological Weyl or nodal line band structure[5,6]. On the other hand, topological Hall effect (THE) is also uncovered in systems with scalar spin chirality[7,8] or topological-nontrivial spin texture[9-11] (skyrmions, typically). These Berry-phase driven phenomena have deeply enriched our understanding of the fundamental physics of magneto transport, especially the Hall effect, as well as offer versatile options to spintronic applications[3,6].

A regular Hall effect is measured during the applied magnetic field (**H**), the electrical current (**I**) and the measured voltage (**V**) are mutually vertical. There exists a so-called planar Hall effect (PHE)[12], where the Hall signal is otherwise obtained in the plane parallel to **H**. PHE is resulted by neither of the mechanism account for AHE, but actually relates to the anisotropic magnetoresistance (AMR)[13,14]. Giant PHE has been reported in the ferromagnetic semiconductor films[15] and recently in the topological Weyl semimetal that correlates with chiral-anomaly[16,17]. However, the nature of PHE makes it exhibit symmetric curves while sweeping **H**, as observed in most cases[15,18], rather than antisymmetric one like a normal Hall effect. In addition, the transverse voltage will vanish when **I** is perpendicular or parallel to **H** according to its phenomenological expression (will be discussed later). But there are several works reported the AHE-like signal in a PHE configuration[19,20], which are currently explained by either a high-order contribution or non-collinear spin structure.

Here we identified a novel planar topological Hall effect (PTHE) in a quasi-two-dimensional van der Waals ferromagnet $Fe_3GeTe_2$[21]. The signal of THE gradually emerges when **H** rotated away from its usual perpendicular direction, and reaches the maximum when H located in the same *ab* plane where the Hall signal is measured. So, we named it as PTHE, to distinguish from the existed THE observed in a normal three-axis configuration. It should be noted that Yihao Wang *et al*[22] have already reported similar-shape THE in $Fe_3GeTe_2$, but the conventional configuration is applied there, making it distinct from the phenomenon observed here. The PTHE was proposed to be originated from the emergent gauge field that associate with the possible topological or chiral spin texture of $Fe_3GeTe_2$. Our results can potentially provide an effective method to detect the in-plane topological spin texture, particular in 2D system and thin films, and further extend the understanding of the Hall effect and may bring new prospective to magneto-transport studies in condensed matter physics.

High-quality single crystal samples were grown through the self-flux method from a mixture of pure elements Fe (99.8%), Ge (99.9999%) and Te (99.99%) with a composition of $Fe_2GeTe_4$[23]. The mixture was then sealed in an evacuated quartz tube and heated to 960℃. The melt was held at 960℃ for 12h, then cooled slowly to 675℃ with a rate of 3℃/h, and finally cooled down to room temperature. Typical size of the single crystals is ~ 2 × 2× 0.1mm, with cleavable layer in the *ab* plane. The crystal structure was identified by X-ray diffraction (XRD, Brucker D8 Advance) with Cu-Kα radiation. Element composition and atomic configuration were examined by energy-dispersive spectroscopy (EDS) in the scanning electron microscope (SEM, FEI Quanta 250F) and high-resolution scanning transmission electron microscopy (STEM, JEOL ARM200F), respectively. The atomic ratio determined by EDS is ~ 3: 1: 2.2, exhibiting slightly off-stoichiometric, which may account for the lowering Curie temperature of our samples comparing to the ones grown by the chemical vapor transport (CVT)[24-26]. Magnetization and transport properties were measured in the superconducting quantum interference device (SQUID) and physical property measurement system (PPMS, Quantum Design) respectively. For transport measurements, the samples were cut into regular rectangle shape and a six-probe method is applied to simultaneously measure the magnetoresistance (MR) and Hall resistivity. The final MR and Hall data were symmetrized to exclude the misalignment of the electrode.

$Fe_3GeTe_2$ is a van der Waals ferromagnet that crystallizes in space group $P6_3/mmc$ with a layered $Fe_3Ge$ substructure sandwiched by two Te layers[21]. As shown in Fig. 1(a), the $Fe_3Ge$ substructure contains two inequivalent $Fe_I$ and $Fe_{II}$ atoms, contributing to a $Fe_{II}$-Ge hexagonal atomic ring layer and two separated triangular lattice $Fe_I$-$Fe_I$ layer. The hexagonal ring was clearly resolved in the high-resolution STEM image in Fig. 1(b), which also confirmed the stacked structure and high quality of our samples. Typical XRD pattern of the as-grown sample in Fig. 1(c) verified the expected single-crystalline nature, with all the reflections along the crystallographic orientation [001].

The temperature dependence of magnetization under zero-field-cooled (ZFC), field-cooled (FC) and field-warming (FW) are measured with $\mu_0H = 0.01T$, both in parallel and perpendicular to the *c* axis, as seen in Fig. 1(d). A ferromagnetic transition is observed at approximately 150K for both directions, similar to that reported for the flux-grown samples[23,27,28]. The ZFC and FC curves show obvious splitting in **H**//*c* below $T_C$ like other typical frustrated ferromagnets[11,29,30], and the difference in magnitude for the two directions indicates the anisotropic character of the low temperature magnetic phase. A slight kink at about 125K below $T_C$ was also observed, consistent with the possible two-stage magnetic ordering transition[26]. The magnetization curve at 5K for both **H**//*c* and **H**//*ab* direction distinctly demonstrates the strong magneto-crystalline anisotropy in $Fe_3GeTe_2$ with the easy axis along the *c* direction. The saturated magnetic moment ($M_S$) is 3.25 $\mu_B$/f. u. for **H**//*c*, also consistent with the reported values[21-23].

The magneto-transport properties for **H** perpendicular to the current plane have been previously investigated[6,22], and a large anomalous Hall current has been identified in the *ab* plane owing to the topological nodal line band structure[6]. In this work, firstly we measured $\rho_{xy}$ and $\rho_{xx}$ in the *ab* plane while gradually rotating the **H** in the *ac* (or *bc*) plane, that is, from the usual *c*-axis to the *ab* plane, as shown in the middle scheme of Fig. 2. We define the angle between the external magnetic field and the normal of the sample plane as $\theta$. When $\theta = 0°$, large anomalous Hall resistivity and Hall angle are reproduced in our samples (supplementary Fig. S1[31]), consistent with previously reported[6]. With $\theta$ increasing from 0° to 90°, we find that $\rho_{xy}$ gradually reduces, while a pronounced cusp-like anomaly arises in the low field region of the curve, as shown in Fig. 2(a). It becomes most distinct for $\theta = 90°$, that is **H**//**I**, which is the reason why the name of PTHE is given. At this configuration, the contribution of the AHE is almost completely excluded. Simultaneously, clear bend at the same filed where PTHE emerges is observed in the MR curves in Fig. 2(b). We proposed that this in-plane Hall and MR anomaly could be resulted by an internal field the pointed in the perpendicular direction, which will be elaborated later.

In the meantime, the full angular dependence from 0-360° of $\rho_{xy}$ and MR in this case are also investigated, as shown in Figs. 2(c, d). In a normal situation, the curve is expected to follow a rough $\cos\theta$ relation, as the AHE scales with the out-of-plane component of **H**[32]. Hence the AHE component at $\theta$ =90 or 270° is expected to vanish, but there remains small value in the figure, which are caused by the misalignment error. It is noted here only at $\mu_0 H = 3T$ that the curve accords with a smooth periodic relation (approaching $\cos\theta$), while at $\mu_0 H = 0.5T$ and 1T, on account of the existence of the large PTHE, it deviates and presents a sharp sign-reverse across the 90 or 270°. Therefore, the PTHE can always be detected in the configuration of **H** // current plane, as observed in several other samples (supplementary Fig. S2[31]). The MR showed two-fold pattern peaking around 90° (270°), with anisotropy of ~ -0.5% (= $(\rho_\parallel - \rho_\perp)/\rho_\perp$) under a field of 3T.

Furthermore, to establish a clearer picture of this abnormal PTHE, we measured the in-plane Hall resistivity $\rho_{xy}$ and MR at various temperatures from 5K to 170K when **H**//**I**, as shown in Figs. 3(a, b). The cusp in $\rho_{xy}$ is being highest at T=100K and persist all temperatures below $T_C$ in the low field region. The MR curves are all negative while bend or upturn emerges at lower temperature. Here the in-plane Hall data was disposed in the same process as that in normal mutual-vertical configuration. The PTHE contribution is denoted as $\rho_{xy}^{PT}$, thus the total Hall resistivity $\rho_{xy} = \mu_0 R_0 H + R_s M + \rho_{xy}^{PT}$, where the first two terms represent the ordinary and anomalous Hall contribution, respectively. As the MR is overall less than 1%, $R_S$ can be simplified as $S_A \rho_{xx}^2$, in which $S_A$ is field independent and $\rho_{xx}$ is the longitudinal resistivity[10,33]. $\rho_{xy}^{PT}$ is thus extracted in combination with the in-plane M-H curves, as showed representatively in Fig. 3(c), which also demonstrates that the peak of $\rho_{xy}$ just appears at the anisotropy field of the M-H curve. By extracting $\rho_{xy}^{PT}$ over the temperature range of 5-170 K, we constitute the phase diagram in Fig. 3(d), by contour mapping in the temperature and magnetic field plane. In general, the PTHE exists from 0.3T to 1.5T, and covers the whole temperature below $T_C$. A maximum value of 2.04 $\mu\Omega \cdot cm$ is obtained at 100K in Fe$_3$GeTe$_2$ single crystal, which is quite large and nearly more than ten times of those observed in systems containing skyrmion phases, such as MnGe[9] (~0.16$\mu\Omega \cdot$ cm), Mn$_3$Ga[34] (~0.26$\mu\Omega \cdot$ cm) and MnNiGa[11] (~0.15$\mu\Omega \cdot$ cm).

Finally, we measured the angular dependence of $\rho_{xy}$ and MR with the field rotated in the *ab* plane at 100K to examine the in-plane anisotropy. The angle between **H** and **I** in the *ab* plane is denoted as $\varphi$. It can be seen in Figs. 4(a, b) that the PTHE can be altered substantially by the change of $\varphi$, while the MR shows slight increase with invariable shape. By sweeping $\varphi$ in the positive (+H) and negative field (-H), we obtain two curves of $\rho_{xy}(\varphi)$ which are actually not superposed (not shown), demonstrating again the existence of large antisymmetric PTHE. The angular dependence of $\rho_{xy}$ extracted by $(\rho_{xy}(+H) - \rho_{xy}(-H))/2$, as shown in the inset of Fig. 4(a), is well-described by the $\sin\varphi$ formulation. Largest magnitude is presented at H=1T, where the PTHE dominates. But it should be noted that the evolution of planar $\rho_{xy}$ with $\varphi$ is not fixed. Our measurements of several samples present varying tendency upon $\varphi$ changing, despite they all follow a rough $\sin(\varphi + x)$ relation (supplementary Fig. S3[31]). Therefore, the largest PTHE is not always observed at **H**//**I**, neither at **H**⊥**I**. Presently we can only attribute the discrepancy to the different state of the surface of our samples, where further investigation is still needed. The so-called PHE, which is obtained by the regular procedure of $(\rho_{xy}(+H) + \rho_{xy}(-H))/2$, is shown in Fig. 4(c). Similar to the reported behavior[18,32], it roughly followed $\rho_{xy}^{PHE} = (\rho_{\parallel} - \rho_{\perp})\sin\varphi\cos\varphi$, where $\rho_{\parallel}$ and $\rho_{\perp}$ are the resistivity when **H** applied parallel or perpendicular to **I**. Hence the variation of $\rho_{xy}^{PHE}$ depends on AMR (expressed as $(\rho_{\parallel} - \rho_{\perp})/\rho_{\perp})$)[35]. Both the magnitude of PHE and AMR is rather small, implying the weak in-plane anisotropy in this uniaxial van der Waals ferromagnet.

Furthermore, from the expression of $\rho_{xy}^{PHE}$, it can be seen that planar $\rho_{xy}^{PHE}$ will vanish when **I** is perpendicular or parallel to **H** ($\varphi = 0\ or\ \pi/2$), and will exhibit symmetric relationship to **H** instead of antisymmetric one like a conventional Hall effect[15,18]. There are some examples where small AHE-like signal is observed[19,20], which are currently explained by a high-order contribution or non-collinear spin structure. But the large presence of PTHE is a phenomenon that first identified. Since there is no external perpendicular component of magnetic field in this configuration, we refer to possible the internal gauge field that account for the extra enhancement of $\rho_{xy}$. On one hand, we notice that the topological (bubble-like) domain structure in the *ab* plane of Fe$_3$GeTe$_2$ has been suspected by the previous MFM and STM measurements[36,37]. The possible in-plane topological spin-texture can give rise to a so-called emergent field[38], which is represented with ***h*** in the schematic inset of Fig. 3(a). The ***h*** behaves just like the real magnetic field to deflect the electrons, thus resulting in the observed Hall signal. Notably, the almost completely exclusion of AHE in the original PTHE makes it an efficient detection method of in-plane skyrmion formation[39]. On the other hand, non-vanishing spin chirality associated with non-coplanar spin structure can induce similar gauge field [1, 22]. As proved by our experiments as well as the literatures [22,24], Fe$_3$GeTe$_2$ is a strong uniaxial magnet with easy axis along the *c* direction (out-of-plane), and meanwhile the Fe atom form a rather frustrated triangular structure (as seen in Fig. 1). Hence during the in-plane magnetization process, the spins of Fe atoms could form a non-coplanar structure that contribute to the ***h*** (also schematically shown in inset of Fig. 3(a)), where further evidence is still needed. These are the two possibilities that can generate an internal field to arouse PTHE in our sample.

In summary, by three-dimensional rotation of the magnetic field with respect to the sample plane, we have observed a large planar topological Hall effect in the hard-magnetized but easily-cleavable *ab* plane in the van der Waals ferromagnet $Fe_3GeTe_2$. Systematic studies on the angular dependence revealed the dynamical evolution of the Hall resistivity and MR which are invariantly measured in the *ab* plane. The PTHE exists in the whole temperature region below $T_C$ and reaches the maximum value of 2.04 $\mu\Omega\cdot cm$ at 1.2T at 100K, which is much larger than those observed in most skyrmion systems. We attribute the origin of this large PTHE to the emergence of a gauge field, which is either generated by the possible topological domain structure of uniaxial $Fe_3GeTe_2$ or the non-coplanar spin structure formed in the process of in-plane magnetization. It should be mentioned that, during the submission of this manuscript, the observation of the topological skyrmion crystal ~~in the *ab* plane~~ has just been~~is~~ reported on the arXiv by T. E. Park *et al*[40], which strongly support one of our presumptions. Therefore, PTHE can be a ~~prominent~~ sensitive ~~way~~ method to detect in-plane skyrmion formation, practically in the natural two-dimensional system and thin films, where the Hall effect on other sides are hardly accessible.

This work is supported by National Natural Science Foundation of China (Grant No. 11604148, 51601092, 11874410), the National Key R&D Program of China (Grant No. 2017YFA0303202), and the Fundamental Research Funds for the Central Universities (Grant No. 30919012108).

# References

1. N. Nagaosa, and Y. Tokura, Phys. Scripta **2012**, 014020 (2012).

2. H. Chen, Q. Niu, and A. H. MacDonald, Phys. Rev. Lett. **112**, 017205 (2014).

3. S. Nakatsuji, N. Kiyohara, and T. Higo, Nature **527**, 212 (2015).

4. A. K. Nayak, J. E. Fischer, Y. Sun, B. Yan, J. Karel, A. C. Komarek, C. Shekhar, N. Kumar, W. Schnelle, J. Kübler, C. Felser, and S. S. P. Parkin, Sci. Adv. **2**, e1501870 (2016).

5. E. Liu, Y. Sun, N. Kumar, L. Muechler, A. Sun, L. Jiao, S. Yang, D. Liu, A. Liang, Q. Xu, J. Kroder, V. Süß, H. Borrmann, C. Shekhar, Z. Wang, C. Xi, W. Wang, W. Schnelle, S. Wirth, Y. Chen, S. T. B. Goennenwein, and C. Felser, Nat. Phys. **14**, 1125 (2018).

6. K. Kim, J. Seo, E. Lee, K.-T. Ko, B. S. Kim, B. G. Jang, J. M. Ok, J. Lee, Y. J. Jo, W. Kang, J.

6. H. Shim, C. Kim, H. W. Yeom, B. I. Min, B.-J. Yang, and J. S. Kim, Nat. Mater. **17**, 794 (2018).

7. Y. Taguchi, Y. Oohara, H. Yoshizawa, N. Nagaosa, and Y. Tokura, Science **291**, 2573 (2001).

8. Y. Machida, S. Nakatsuji, Y. Maeno, T. Tayama, T. Sakakibara, and S. Onoda, Phys. Rev. Lett. **98**, 057203 (2007).

9. N. Kanazawa, Y. Onose, T. Arima, D. Okuyama, K. Ohoyama, S. Wakimoto, K. Kakurai, S. Ishiwata, and Y. Tokura, Phys. Rev. Lett. **106**, 156603 (2011).

10. S. X. Huang, and C. L. Chien. Phys. Rev. Lett. **108**, 267201 (2012).

11. W. Wang, Y. Zhang, G. Xu, L. Peng, B. Ding, Y. Wang, Z. Hou, X. Zhang, X. Li, E. Liu, S. Wang, J. Cai, F. Wang, J. Li, F. Hu, G. Wu, B. Shen, and X. -X. Zhang, Adv. Mater. **28**, 6887 (2016).

12. J. P. Pan, in Solid State Physics, edited by F. Seitz and D. Turnbull (Academic, New York, 1957), Vol. 5, pp. 1–96.

13. Z. Q. Lu, G. Pan, and W. Y. Lai, J. Appl. Phys. **90**, 1414 (2001).

14. K. M. Seemann, F. Freimuth, H. Zhang, S. Blügel, Y. Mokrousov, D. E. Bürgler, and C. M. Schneiderl, Phys. Rev. Lett. **107**, 086603 (2011).

15. H. X. Tang, R. K. Kawakami, D. D. Awschalom, and M. L. Roukes, Phys. Rev. Lett. **90**, 107201 (2003).

16. A. A. Burkov, Giant planar Hall effect in topological metals. Phys. Rev. B **96**, 041110 (2017).

17. S. Nandy, Girish Sharma, A. Taraphder, and S. Tewari, Phys. Rev. Lett. **119**, 176804 (2017).

18. H. Li, H.-W. Wang, H. He, J. Wang, and S.-Q. Shen, Phys. Rev. B **97**, 201110 (2018).

19. P. K. Muduli, K.-J. Friedland, J. Herfort, H.-P. Schönherr, and K. H. Ploog, Phys. Rev. B **72**, 104430 (2005).

20. Y. You, X. Chen, X. Zhou, Y. Gu, R. Zhang, F. Pan, and C. Song, Adv. Electron. Mater. **5**, 1800818 (2019).

21. H.-J. Deiseroth, K. Aleksandrov, C. Reiner, L. Kienle, and R. K. Kremer, Eur. J. Inorg. Chem. **2006**, 1561 (2006).

22. Y. Wang, C. Xian, J. Wang, B. Liu, L. Ling, L. Zhang, L. Cao, Z. Qu, and Y. Xiong, Phys. Rev. B **96**, 134428 (2017).

23. A. F. May, S. Calder, C. Cantoni, H. Cao, and M. A. McGuire, Phys. Rev. B **93**, 014411

(2016).

24. B. Chen, J. Yang, H. Wang, M. Imai, H. Ohta, C. Michioka, K. Yoshimura, and M. Fang, J. Phys. Soc. Jpn. **82**, 124711 (2013).

25. V. Y. Verchenko, A. A. Tsirlin, A. V. Sobolev, I. A. Presniakov, and A. V. Shevelkov, Inorg. Chem. **54**, 8598 (2015).

26. J. Yi, H. Zhuang, Q. Zou, Z. Wu, G. Cao, S. Tang, S. A. Calder, P. R. C. Kent, D. Mandrus, and Z. Gai. 2D Mater. **4**, 011005 (2016).

27. Y. Liu, E. Stavitski, K. Attenkofer, and C. Petrovic, Phys. Rev. B **97**, 165415 (2018).

28. Y. Liu, V. N. Ivanovski, and C. Petrovic, Phys. Rev. B **96**, 144429 (2017).

29. Z. Hou, W. Ren, B. Ding, G. Xu, Y. Wang, B. Yang, Q. Zhang, Y. Zhang, E. Liu, F. Xu, W. Wang, G. Wu, X. Zhang, B. Shen, and Z. Zhang, Adv. Mater. **29**, 1701144 (2017).

30. Z. Hou, Q. Zhang, G. Xu, S. Zhang, C. Gong, B. Ding, H. Li, F. Xu, Y. Yao, E. Liu, G. Wu, X.-X. Zhang, and W. Wang, ACS Nano **13**, 922 (2019).

31. See Supplemental Material at [*URL will be inserted by publisher*]. Fig. S1 shows the Hall resistivity $\rho_{xy}$ and MR at variable temperature when the magnetic field is applied along the *c* direction. Fig. S2 shows the magnetic field dependence of Hall resistivity $\rho_{xy}$ and longitudinal resistivity with **H** // **I** at various temperature for another two samples. Fig. S3 shows the angular dependence of $\rho_{xy}$ of three other samples (S5-S7) at 100K under the field of 1T, 3T and 5T.

32. N. Kumar, S. N. Guin, C. Felser, and C. Shekhar, Phys. Rev. B **98**, 041103 (2018).

33. Y. Li, N. Kanazawa, X. Z. Yu, A. Tsukazaki, M. Kawasaki, M. Ichikawa, X. F. Jin, F. Kagawa, and Y. Tokura, Phys. Rev. Lett. **110**, 117202 (2013).

34. P. Sergelius, J. Gooth, S. Bäßler, R. Zierold, C. Wiegand, A. Niemann, H. Reith, C. Shekhar, C. Felser, B. Yan and K. Nielsch, Sci. Rep. **6**, 33859 (2016).

35. D. Thompson, L. Romankiw, and A. Mayadas. IEEE Trans. Magn. **11**, 1039 (1975).

36. N. León-Brito, E. D. Bauer, F. Ronning, J. D. Thompson, and R. Movshovich, J. Appl. Phys. **120**, 083903 (2016).

37. G. D. Nguyen, J. Lee, T. Berlijn, Q. Zou, S. M. Hus, J. Park, Z. Gai, C. Lee, and A. P. Li, Phys. Rev. B **97**, 014425 (2018).

38. N. Nagaosa and Y. Tokura. Nat. Nanotech. **8**, 899 (2013).

39. T. Yokouchi, N. Kanazawa, A. Tsukazaki, Y. Kozuka, A. Kikkawa, Y. Taguchi, M. Kawasaki,

M. Ichikawa, F. Kagawa, and Y. Tokura, J. Phys. Soc. Jpn. **84**, 104708 (2015).

40. T. E. Park, L. Peng, X. Zhang, S. J. Kim, K. M. Song, K. Kim, M. Weigand, G. Schütz, S. Finizio, J. Raabe, J. Xia, Y. Zhou, M. Ezawa, X. Liu, J. Chang, H. C. Koo, Y. D. Kim, X. Yu, S. Woo, arxiv: 1907.01425.

# Figures and captions

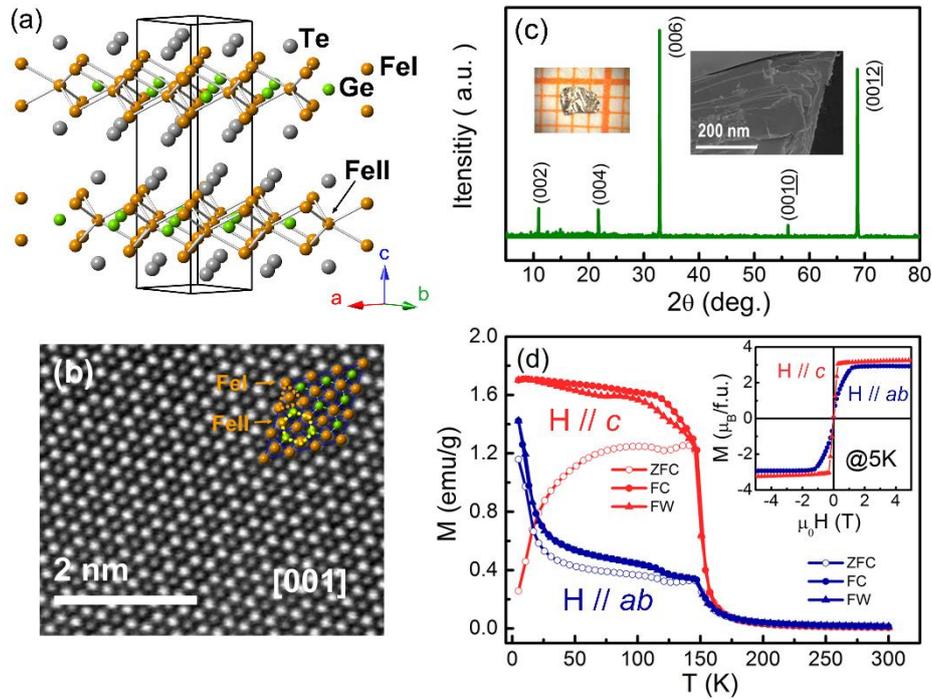

FIG. 1. Structure and magnetic properties of the single crystal $Fe_3GeTe_2$. (a) Schematic of the hexagonal structure of $Fe_3GeTe_2$ (space group *P6_3/mmc*). (b) High-resolution STEM image taken from the [001] axis. The inset shows the arrangement of the triangular structure of $Fe_I$ atoms and the hexagonal ring of $Fe_{II}$-Ge layer. (c) XRD pattern for the as-grown slice of $Fe_3GeTe_2$ single crystal. The optical photograph and SEM image in the inset shows the typical size and stepped-like appearance of the sample. (d) Temperature dependence of the ZFC, FC and FW magnetization measured at $\mu_0 H = 0.01$T for **H**//*ab* and **H**//*c*. Inset shows the anisotropic M-H curve at T = 5K.

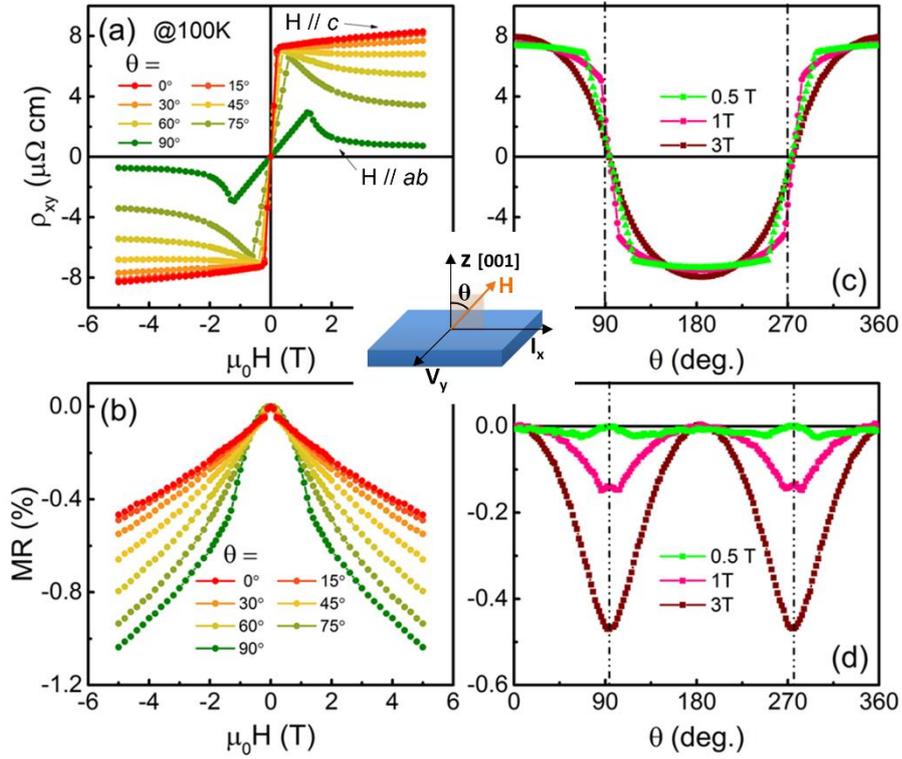

FIG. 2. Separate (a) $\rho_{xy}(H)$ and (b) MR curves measured in the *ab* plane with **H** in direction of $\theta = 0°, 15°, 30°, 45°, 60°, 75°$ and $90°$ at 100K. The angular dependence of (c) $\rho_{xy}$ and (d) MR at 100K under the field of 0.5T, 1T and 3T. In the middle it shows the schematic of the measurement configuration.

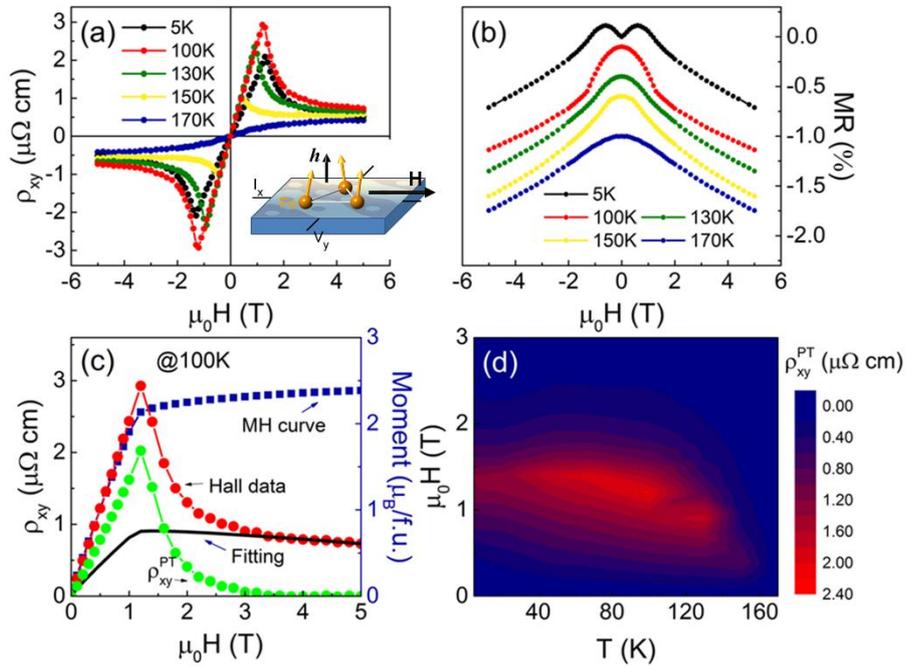

FIG. 3. In-plane (a) $\rho_{xy}$ and (b) MR of $Fe_3GeTe_2$ single crystal at various temperatures for **H**//**I**. Inset is the configuration of the measurement, as well as schematic of the possible origin that generated an inside gauge filed of *h*. (c) The representative $\rho_{xy}$ curve (red) and MH curve (blue) measured at 100K, calculated $R_0B + S_A\rho_{xx}^2M$ fitting curve (black) and extracted $\rho_{xy}^{PT}$ curve (green). (d) The contour mapping of extracted $\rho_{xy}^{PT}$ as a function of the external magnetic field and temperature.

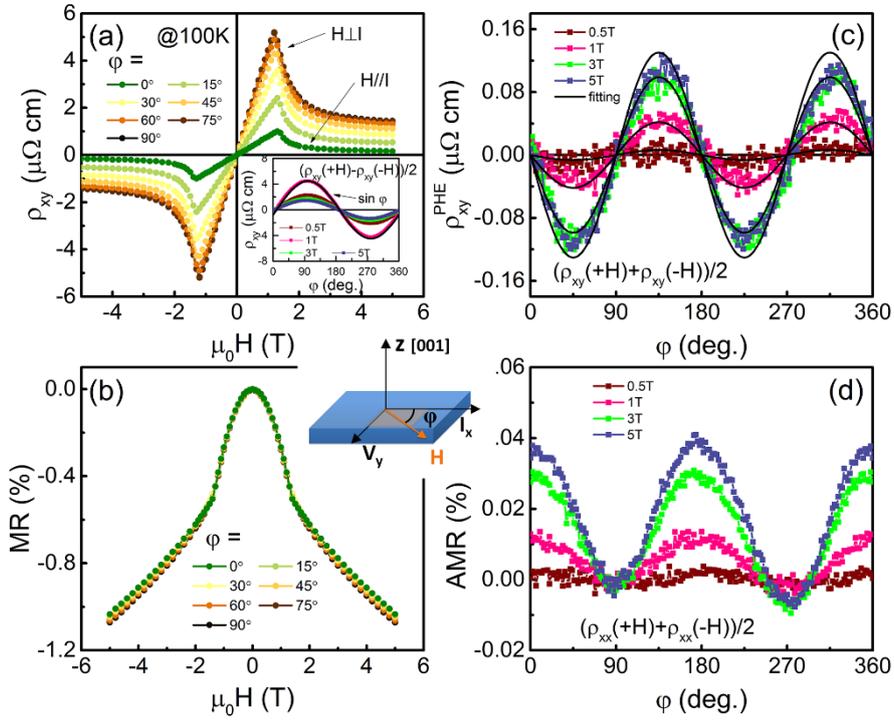

FIG. 4. Magnetic field dependence of planar (a) $\rho_{xy}$ and (b) MR at $\varphi = 0°, 15°, 30°, 45°, 60°, 75°$ and $90°$ at 100K. The inset in (a) is the angular dependence of $\rho_{xy}$ by subtracting the original curve in positive and negative field $((\rho_{xy}(+H) - \rho_{xy}(-H))/2)$. The angular dependence of the (c) $\rho_{xy}^{PHE}$, obtained by $(\rho_{xy}(+H) + \rho_{xy}(-H))/2$ and (d) AMR at 100K under the field of 0.5T, 1T, 3T and 5T.

**Supplementary Materials**

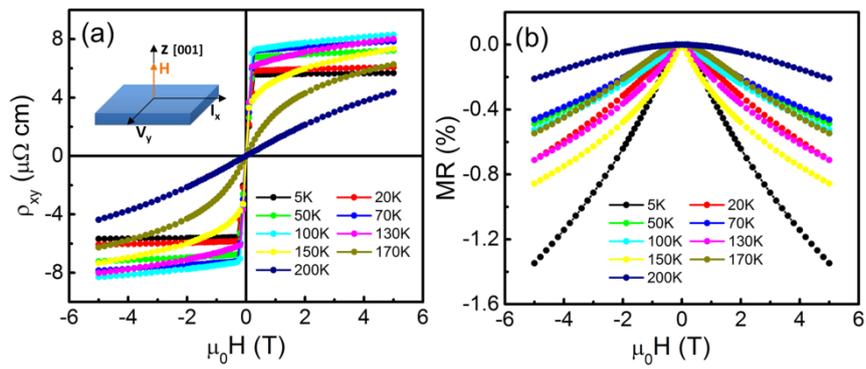

FIG. S1. Variable temperature (a) Hall resistivity $\rho_{xy}$ and (b) MR of the same sample as measured in Fig. 2 and Fig. 3, when the magnetic field is applied along the *c* direction.

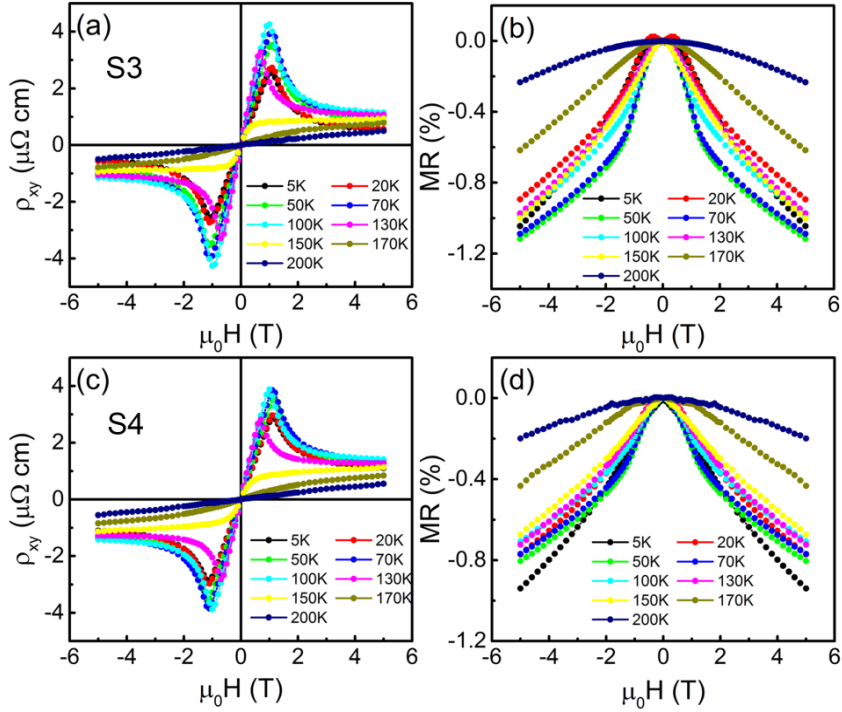

FIG. S2. Magnetic field dependence of (a, c) Hall resistivity $\rho_{xy}$ and (b, d) longitudinal resistivity with **H** // **I** at various temperature for another two samples labeled as S3 and S4.

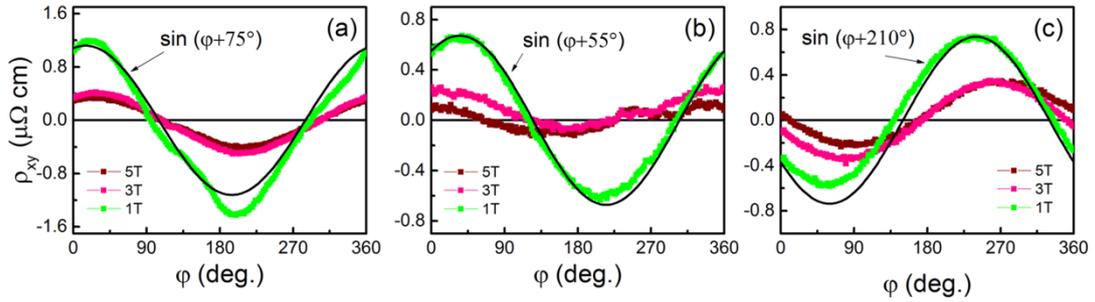

FIG. S3. Angular dependence of $\rho_{xy}$, by taking the difference of the original curve for positive and negative field $((\rho_{xy}(+H) - \rho_{xy}(-H))/2)$, of three other samples (S5-S7) at 100K under the field of 1T, 3T and 5T.